# Dark Matter as a consequence of electric charge non-conservation - will it remain "dark" forever?

Richard M. Weiner

Laboratoire de Physique Théorique, Univ. Paris-Sud, Orsay, France and

Physics Department, University of Marburg, Germany[x)]

It is conjectured that dark matter (DM) was produced before inflation from neutral particles present after the Big Bang and survived inflation due to a specific coupling with gravitation, while the charged particles existing after the Big Bang disappeared during inflation in a process of charge non-conservation. Ordinary matter was produced at a later stage at a lower temperature following a symmetry restoring phase transition. In this way the non-luminous character of dark matter and the existence of two types of matter, ordinary and dark, get a natural explanation. Because of the high temperatures preceding inflation, the masses of particles produced during that time are too large to be detected by conventional particle physics methods and possibly will never be detected.

Dark Matter (DM) constitutes one of the most challenging, unsolved problems of cosmology and of physics in general (for more recent reviews cf. e.g. refs. [1]): although it is responsible for about 85% of the total mass density of the Universe, the mechanism of its creation, its constituents and the reason for its main property, its darkness, are unknown. At present the most popular attempts of explanations of DM are based on the assumption that its constituents are weakly interacting massive particles (WIMP -s), which are electrically neutral. However why these constituents have not been detected in the laboratory, despite the fact that other types of neutral weakly interacting particles like neutrinos have been, is unknown. Moreover within the presently accepted lore of cosmology - the Big Bang followed by inflation and then by standard model physics - the question which is at the heart of the dark matter mystery "why does the Universe contain two types of matter, one 'ordinary' and accessible to optical devices and another one, 'dark', remains unanswered. In other words, is the existence of two types of matter, one dark and one ordinary, a natural consequence of the evolution of the Universe, as we believe to know it at present?

---

[x)] Email address: weiner@staff.uni-marburg.de

One of the aims of this note is to suggest a positive answer to this question and to show that the creation of dark matter might cease to be a mystery once we give up the assumption, made so far in all approaches to DM, that one of the laws of physics, that of conservation of electric charge, which is valid in our universe at present, was also valid immediately after the Big Bang. In particular we assume that dark matter was produced as a consequence of the fact that the charged particles created immediately after the Big Bang lost their electric charge during inflation. Charge conservation was established at a later stage. This assumption is also used to explain why DM has not yet been "seen" and to suggest that it possibly will never be.

We start by considering the Universe after the Big Bang. What appears quite well established is the fact that it underwent a period of inflation.
Without losing generality it can be assumed that before this happened it consisted of charged and neutral particles. In accordance with what is known at present about inflation charged particles disappeared during inflation. This observation due initially to Guth [2] has remained unchallenged by the subsequent formulations of inflation. From the point of view of the early Universe this represents a process of global charge non-conservation. On top of that we assume that due to a specific coupling with the gravitational field to be described below neutral particles survived inflation. These "early" neutral particles constitute what appears at present dark matter. For the sake of concreteness (cf. below) we will assume that these particles are thermal relics. Ordinary matter was produced after inflation in a charge symmetry restoring phase transition and consists of charged and "late" neutral particles, which besides the absence of the specific gravitational coupling mentioned above, differ from the "early" ones, among other things, by their mass.
This explanation of the existence of dark and ordinary matter constitutes in a certain sense a modest attempt to follow Einstein's advice to try to "understand not only how nature works, but to understand why nature is the way it is".
Since according to our main postulate dark matter is produced before inflation which takes place at a temperature
$$T_{DM} \sim T_{infl} \approx 10^{-3}\text{-}10^{-4} M_{Planck} \; , \qquad\qquad\qquad\qquad (1)$$

the average mass $M_{DM}$ of its constituents is by orders of magnitude larger than that of ordinary matter $M_{ord}$, which were produced later at temperatures

$$T_{ord} < T_{infl} . \qquad (2)$$

This is the reason why DM constituents have not been detected experimentally in the laboratory, either in accelerator or in cosmic rays physics.

That the assumption of charge non-conservation is not so shocking as it might appear at a first look can be realized by reminding that charge conservation, like most other conservation laws, is, according to Noether's theorem, a consequence of a symmetry of the Lagrangian, in this case the electromagnetic gauge symmetry. On the other hand, whether a given solution of the equations of motion of a Lagrangian exhibits a given symmetry or not depends on external circumstances like temperature, matter densities or external fields and on details of the Lagrangian. In particular, the fact that most known symmetries are conserved at high temperatures led by analogy to the assumption that this is also valid for the very early periods of the evolution of the Universe, so that most conservation laws known at present were also valid immediately after the Big Bang.

However ferromagnets like the Rochelle salt are a well known counter example of the observation that high temperatures lead to more symmetry [3]. What is even more relevant for the present approach is the fact that for global symmetries the breaking of symmetry at high temperatures is quite a general phenomenon, with possible implications for the early stages of the Universe expansion [4]. Actually non- conservation of electric charge at high temperatures was proposed already a long time ago [5] in order to explain the absence of magnetic monopoles. Similarly the assumption of symmetry violation at high temperatures was used to explain baryogenesis [6].

In the present approach we assume that the gauge symmetry associated with electric charge was spontaneously broken after the Big Bang and subsequently restored at lower temperatures, during the expansion of the Universe [7]. That happened through a phase transition at a critical temperature $T_{crit} \geq T_{ord}$.

This assumption shifts the DM creation to the very early stages of the Universe evolution, when gravitational effects are still important. And although the quantum field theory of gravitation is non-renormalizable, almost everything we know today about the beginning of the Universe is based, besides the theory of general relativity, on classical gravitation. We assume that this applies also for the following considerations, the more so that quantum effects are most probably not important at high temperatures [8].

The question then arises how dark matter created before inflation survived the exponential expansion characteristic for inflation, which is expected to wash out any conserved quantity. Here gravitation comes into play through the "mimetic" mechanism of Chamseddine and Mukhanov [9], who described dark matter by a scalar field $\Phi$ coupled to the inflaton field $\varphi$ through a term of the form $\Phi F(\varphi)$, where F is a slow function of $\varphi$.

These authors prove that at the end of the exponential expansion representing inflation

$$a \equiv H^{-1} \exp(Ht), \tag{3}$$

where a defines the flat metric

$$ds^2 = dt^2 - a^2(t)\delta_{ik}dx^i dx^k, \tag{4}$$

and H is the Hubble constant, the diference between the trace G of the Einstein tensor $G^{\mu\nu} = R^{\mu\nu} - \frac{1}{2}Rg^{\mu\nu}$ and the trace T of the energy momentum tensor of matter $T^{\mu\nu}$ can be approximated by

$$G - T \approx -F(\varphi)/3H \neq 0, \tag{5}$$

which means that dark matter as defined above survives inflation.

Further development of this approach might include e.g. the use of the experimentally observed relic density of dark matter in the determination of yet unknown aspects and parameters of the mimetic formalism, a topic of current interest (cf. e.g.[10]).

To summarize: the existence of dark matter can be considered as a consequence of the breakdown in the early stages of Universe expansion of a fundamental principle of particle physics, electric charge conservation. This conclusion emerges independently of any more detailed quantitative considerations.

The existence of particles with masses up to the inflation temperature $T_{infl}$ could contribute to solving the hierarchy problem, i.e. filling the gap between the Higgs mass characteristic for the electroweak interaction and the Planck mass characteristic for gravitation.

The most far reaching *experimental* consequence of the above considerations is the fact that because of the high mass of its constituents dark matter may remain "dark" forever in the sense that its constituents will never be observed in the laboratory, although it represents one of the strongest signals of the early Universe.

Acknowledgements: The continuous advice and interest of Masud Chaichian in this work is gratefully acknowledged. I am also much indebted to Maxim Khlopov for important suggestions, to Paul Langacker, Markku Oksanen and Cristian Armendariz-Picon for instructive observations and to Tom Kibble and Archil Kobakhidze for encouraging comments.

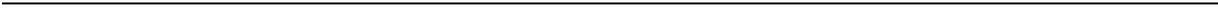